\documentclass[aps,twocolumn,prl,floatfix,superscriptaddress,showpacs,longbilbiography]{revtex4-1}

\usepackage[english]{babel}
\usepackage[utf8]{inputenc}
\usepackage[T1]{fontenc}

\usepackage{amssymb}
\usepackage{amsmath}
\usepackage{amsbsy}
\usepackage{bm}
\usepackage{graphicx}
\usepackage{float}
\usepackage{braket}

\usepackage{color,soul}
\usepackage{wasysym}
\usepackage{mathrsfs}
\usepackage{times}

\usepackage[dvipsnames,svgnames,table]{xcolor}
\usepackage{hyperref}
\usepackage{fancyhdr}
\usepackage{float}
\usepackage[english]{babel}
\usepackage[caption=false]{subfig}
\usepackage{braket}
\hypersetup{
pdfstartview={FitH},% fits the width of the page to the window
colorlinks=true,    % false: boxed links; true: colored links
linkcolor=NavyBlue, % color of internal links
citecolor=Maroon,   % color of links to bibliography
filecolor=NavyBlue, % color of file links
urlcolor=NavyBlue   % color of external links
}

\makeatletter
\def\bbl@set@language#1{%
  \edef\languagename{%
    \ifnum\escapechar=\expandafter`\string#1\@empty
    \else\string#1\@empty\fi}%
  \@ifundefined{babel@language@alias@\languagename}{}{%
    \edef\languagename{\@nameuse{babel@language@alias@\languagename}}%
  }%
  \select@language{\languagename}%
  \expandafter\ifx\csname date\languagename\endcsname\relax\else
    \if@filesw
      \protected@write\@auxout{}{\string\select@language{\languagename}}%
      \bbl@for\bbl@tempa\BabelContentsFiles{%
        \addtocontents{\bbl@tempa}{\xstring\select@language{\languagename}}}%
      \bbl@usehooks{write}{}%
    \fi
  \fi}
\newcommand{\DeclareLanguageAlias}[2]{%
  \global\@namedef{babel@language@alias@#1}{#2}%
}
\makeatother
\DeclareLanguageAlias{en}{english}

\begin{document}
\title{Topological magnonics in the two-dimensional van der Waals magnet $\text{CrI}_3$}
\author{Esteban Aguilera}
\affiliation{Departamento de F\'isica, FCFM, Universidad de Chile, Santiago, Chile.}
\author{R. Jaeschke-Ubiergo}
\affiliation{Departamento de F\'isica, FCFM, Universidad de Chile, Santiago, Chile.}
\author{N. Vidal-Silva}
\affiliation{Departamento de F\'isica, FCFM, Universidad de Chile, Santiago, Chile.}
\author{L. E. F. Foa Torres}
\affiliation{Departamento de F\'isica, FCFM, Universidad de Chile, Santiago, Chile.}
\author{A. S. Nunez}
\affiliation{Departamento de F\'isica, FCFM, Universidad de Chile, Santiago, Chile.}
\affiliation{CEDENNA, Universidad de Santiago de Chile, Avda. Ecuador 3493, Santiago, Chile.}

\begin{abstract}
In this article, we calculate the magnon spectrum of Kitaev-Heisenberg magnets. This model has been recently proposed as a spin Hamiltonian to model $\text{CrI}_3$ and other two-dimensional magnets. It is a minimal spin Hamiltonian that includes a contribution stemming from a  Heisenberg, isotropic exchange, and a contribution arising from a Kitaev interaction, anisotropic and frustrated exchange. Our calculations reveal the topological nature of the magnons and a gap that opens at the $K$ and $K^\prime$ points. These topological properties give rise to effects such as thermal Hall effect. In addition to the bulk properties, we calculate the magnon spectrum of nanoribbons illustrating the corresponding edge states.
\end{abstract}

\maketitle

{\em Introduction.-} The graphene revolution led the way into the world of two-dimensional materials~\cite{novoselov_electric_2004}, with properties that baffled the usual behavior found in their three-dimensional counterparts. The path was followed with enthusiasm by early practitioners and has fruitfully rewarded them with a plethora of 2D-materials and van der Waals heterostructures~\cite{novoselov_2d_2016} with groundbreaking properties. The list of discoveries grows steadily and includes a variety of semiconductors~\cite{castellanos-gomez_why_2016}, superconductors~\cite{Lu, Ugeda} and ferroelectrics~\cite{Chang}. Spintronic devices~\cite{roche_graphene_2015} are also intensively targeted~\cite{benitez_tunable_2020} as one of the most promising applications. In 2017, new forms of 2D materials were reported to display ferromagnetism,  a state of matter elusive in the two-dimensional realm until then~\cite{Huang, Gong}. In particular, a van der Waals material, single layer $\text{CrI}_3$, was reported to display ferromagnetism~\cite{Huang} under 45K. To overcome 2D thermal fluctuations that would otherwise render its magnetization unstable~\cite{MerminWagner}, $\text{CrI}_3$ relies heavily upon several forms of anisotropy~\cite{Lado_2017}.

In this work, we report on the collective behavior of the magnetic degrees of freedom of two dimensional magnets such as $\text{CrI}_3$ in the form of magnons. Magnons are quantized low energy excitations of the magnetization field~\cite{Auerbach}. Their control and manipulation might lead to novel applications in the field of magnon spintronics~\cite{Kruglyak_2010, Chumak}. As we detail below, these excitations in $\text{CrI}_3$ seem to defy the standard wisdom in magnetism and display unusual behavior with potential applications in several areas such as quantum computing and spintronics.

 The reason for these unusual properties is that $\text{CrI}_3$, as recently proposed~\cite{Lee, Xu}, is described by a Kitaev's interaction. Since the material is essentially composed of an honeycomb lattice of edge sharing octahedra, it is natural to expect similarities with layered $Na_2IrO_3$ and $a-RuCl_3$ well known for its behavior as a spin liquid dominated by a Kitaev Hamiltonian~\cite{Jackeli}. This interaction is an anisotropic form of frustrated exchange that, when acting alone, unleashes a formidable gallery of topologically protected magnetic excitations, such as anyons and Majorana excitations~\cite{Kitaev, Hermanns}. Like other systems that have been proposed as implementation of topological magnonics~\cite{Murakami, Owerre_2016, Rold_n_Molina_2016,Hidalgo},  the current proposal offers a way into controllable excitations with great potential in the context of magnonic devices.

{\em Basic Model.-} The magnetic degrees of freedom in the $\text{CrI}_3$ ferromagnet can be modeled using the Heisenberg-Kitaev model. Chromium sites form a magnetic honeycomb lattice, with magnetic moment $S=3/2$. The Hamiltonian consists in the usual isotropic Heisenberg exchange, plus an anisotropic contribution that comes from Kitaev model. The magnetic Hamiltonian takes the form

\begin{equation}
    H = -\sum_{< i, j >} \left(J \mathbf{S}_i \cdot \mathbf{S}_j + K S_{i}^{\gamma}  S_{j}^{\gamma}\right)- \sum_i A (S_i^z)^2.
    \label{Kitaev_H}
\end{equation}
 Here the first summation runs over nearest neighbours, and we define $S_i^{\gamma}\equiv \mathbf{S}_i\cdot \boldsymbol{\hat{\gamma}}$, as the component of the magnetic moment in the $\mathbf{\hat{\gamma}}$ direction. These directions depend on the link, so $\boldsymbol{\hat{\gamma}}$ should be understood in \eqref{Kitaev_H} as an abbreviation of  $\boldsymbol{\hat{\gamma}}_{ij}$. $J$ and $K$ are the Heisenberg and Kitaev exchange constants respectively. We also include an easy-axis anisotropy of magnitude $A$.

\begin{figure}
    \centering
    \includegraphics[scale=0.30]{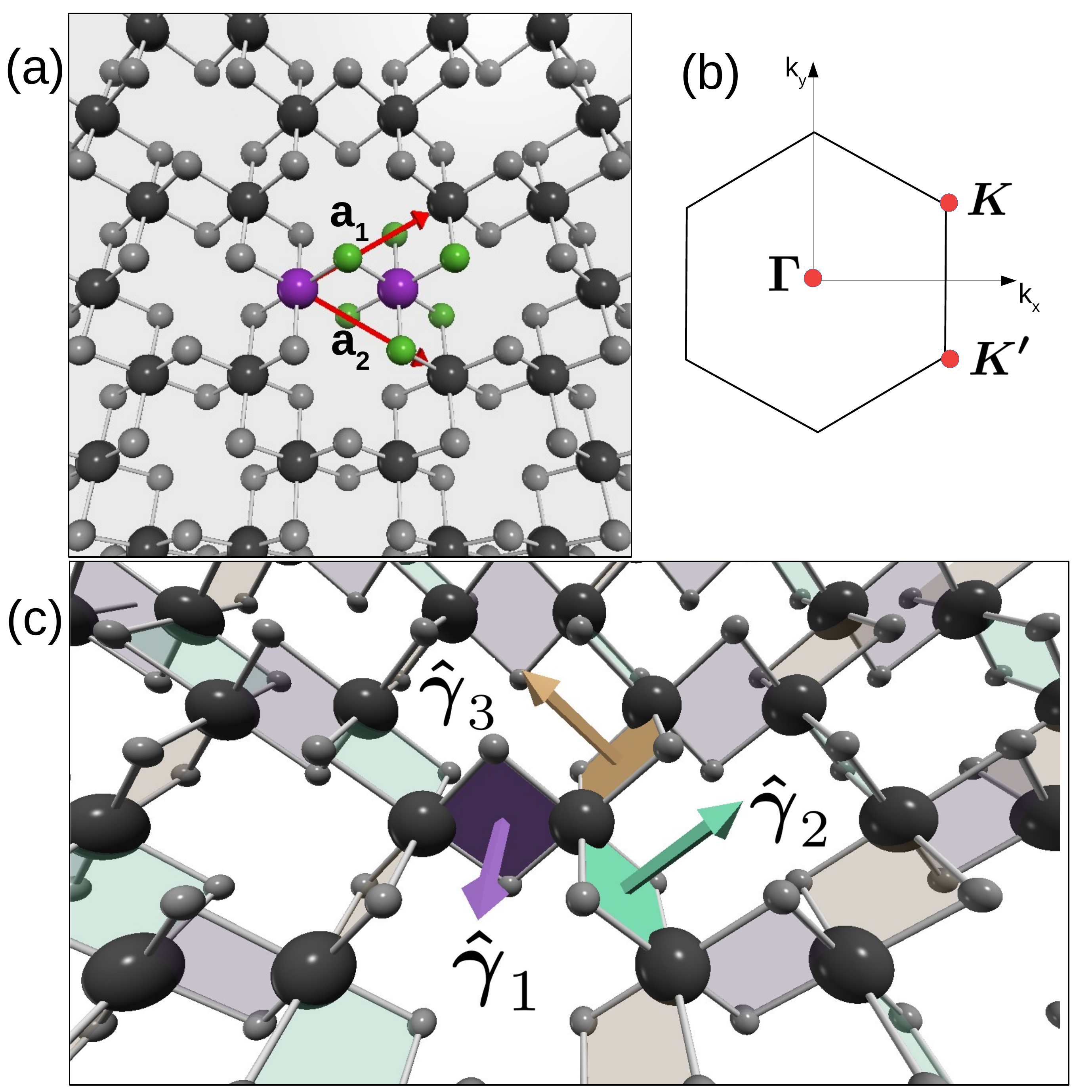}
    \caption{(a) Top view of $\text{CrI}_3$ monolayer. The atoms in the unit cell are highlighted in color, big purple spheres represent Chromium sites, and small green spheres represent Iodine sites. Lattices vectors $\mathbf{a_1}=a_0(\frac{\sqrt{3}}{2},\frac{1}{2},0)$ and $\mathbf{a_2}=a_0(\frac{\sqrt{3}}{2},-\frac{1}{2},0)$ were drawn with red arrows, $a_0$ is the lattice constant. (b) First Brillouin zone, with special symmetry points $\boldsymbol{\Gamma}$, $\boldsymbol{K}$ and $\boldsymbol{K'}$.(c) A view of the $\text{CrI}_3$ monolayer in perspective, with the plaquettes $Cr_2I_2$ colored according to their normal vectors $\boldsymbol{\hat{\gamma}_1}$,$\boldsymbol{\hat{\gamma}_2}$ and $\boldsymbol{\hat{\gamma}_3}$. Three plaquettes in the unit cell are highlighted, and normal vectors form an orthonormal basis.
    }
    \label{fig1}
\end{figure}

We are considering nearest neighbours in a honeycomb lattice, so we have three kinds of links on each unit cell. Fig.~\ref{fig1} shows the links and the respective $\boldsymbol{\hat{\gamma}}_a$ directions. Note that each $\boldsymbol{\hat{\gamma}}_a$ points normal to the $Cr_2I_2$ plaquette that contains the link $\mathbf{\hat{l}}_a$. The explicit form of $\boldsymbol{\hat{\gamma}}_a$ vectors  in the basis $xyz$ is
$
    \boldsymbol{\hat{\gamma}}_1 = \left(0, \frac{-\sqrt{2}}{\sqrt{3}}, \frac{1}{\sqrt{3}}\right),$ $
    \boldsymbol{\hat{\gamma}}_2 = \left(\frac{1}{\sqrt{2}}, \frac{1}{\sqrt{6}}, \frac{1}{\sqrt{3}}\right)$ and
    $\boldsymbol{\hat{\gamma}}_3 =\boldsymbol{\hat{\gamma}}_1\times \boldsymbol{\hat{\gamma}}_2$.
Note that $\{ \boldsymbol{\hat{\gamma}}_a\}$ vectors form an orthonormal basis oriented as shown in fig. \ref{fig1}. The orientation of this triad and the presence of the additional anisotropy make this model different from the one described in \cite{Gohlke, Joshi}.

Both the Heisenberg and Kitaev contributions to exchange can be put together using a exchange matrix $\mathcal{J}_{ij}$, in the form
$-\sum_{<i, j>} \mathbf{S}_i \cdot \mathcal{J}_{ij} \cdot \mathbf{S}_j$,
where $\mathcal{J}_{ij}$ can take, depending on the link, one of three different forms $\mathcal{J}_1$, $\mathcal{J}_2$ or $\mathcal{J}_3$. The matrix  $\mathcal{J}_a$ takes the form:
$ \mathcal{J}_{a} = J \mathbf{1} + K \gamma_a\otimes\gamma_a^. $
Note that $\mathcal{J}_a$ remains invariant under the transformation $\boldsymbol{\hat{\gamma}}_a \rightarrow -\boldsymbol{\hat{\gamma}}_a$.

To obtain the Hamiltonian for magnons, we use the Holstein-Primakoff's \cite{HolsteinPrimakoff} transformation: $
        S_{i\mu}^{(x)} = \sqrt{\frac{S}{2}} \left(\psi^{\dagger}_{i\mu} + \psi_{i\mu}\right),$
        $S_{i\mu}^{(y)} = i\sqrt{\frac{S}{2}} \left(\psi^{\dagger}_{i\mu} - \psi_{i\mu}\right)$,
        $S_{i\mu}^{(z)} = S - \psi^{\dagger}_{i\mu} \psi_{i\mu}$,
        where $\mu\in \{A,B\}$ indexes the two lattices conforming the bipartite honeycomb array of $Cr$ atoms.

When replaced in the Hamiltonian and reduced to quadratic terms, we obtain a Hamiltonian in terms of $
    \boldsymbol{\Psi_k} = (
    \psi_{A\mathbf{k}},
    \psi_{B\mathbf{k}},
    \psi^\dagger_{A -\mathbf{k}},
    \psi^\dagger_{B-\mathbf{k}})^t$, in the form $\mathcal{H}=\sum_{\mathbf{k}}\boldsymbol{\Psi_k}^\dagger \mathbf{H}_k \boldsymbol{\Psi_k}$ where

    \begin{equation}\label{Kitaev_Hk}
    \mathbf{H}_{k} = \left(\begin{aligned}
    \varepsilon &&\alpha_k && 0 &&\beta_k \\
    \alpha_k^{*} && \varepsilon  && \beta_{-k} && 0\\
    0 && \beta^{*}_{-k} && \varepsilon && \alpha_k \\
    \beta^{*}_{k} && 0 && \alpha^{*}_k && \varepsilon
    \end{aligned}\right)
\end{equation}
 In the above expression we have defined $\varepsilon = S(3J + K + 2A)$,
$
    \alpha_k = \sum_{a} \alpha_a e^{i\mathbf{k}\cdot \boldsymbol{\delta}_a}
$
, and
$
    \beta_k = \sum_{a} \beta_a e^{i\mathbf{k}\cdot \boldsymbol{\delta}_a}
$, where we have defined the following quantities: $
    \alpha_a = -S(\mathcal{J}_a^{xx}+\mathcal{J}_a^{yy})/2$,
    $\beta_a = -S(\mathcal{J}_a^{xx} -\mathcal{J}_a^{yy})/2-i S\mathcal{J}_a^{xy}$. Also $\boldsymbol{\delta}_a$ vectors are expressed in terms of lattice vectors as: $\boldsymbol{\delta}_1=0$, $\boldsymbol{\delta}_2=\mathbf{a_1}$ and $\boldsymbol{\delta}_3=\mathbf{a_2}$.

A Bogoliubov transformation~\cite{Colpa} allows us to obtain the eigen-energies and their corresponding eigen-states.  In terms of $\epsilon_0=S(3J+K)$,  $  \mathcal{K} = KS/\epsilon_0$ and $\mathcal{A}=2AS/\epsilon_0$,  the eigenvalues are found to be:

\begin{equation}
\epsilon_{\pm}^{2}(\mathbf{k}) = \frac{\epsilon_0^2}{9}\left(f(\mathbf{k}) \pm \sqrt{g(\mathbf{k})}\right)
\end{equation}
where the functions $f=f_0 + f_{\mathcal{A}} + f_{\mathcal{K}}$ and $g=g_0 + g_{\mathcal{A}} + g_{\mathcal{K}}$  are defined in the supplementary material, separating the contributions of Kitaev and Anisotropy terms, in such a way $f_{\mathcal{A},\mathcal{K}}$ and $g_{\mathcal{A}, \mathcal{K}}$ are zero when anisotropy or Kitaev terms are neglected.

\begin{figure}[H]
        \includegraphics[width=0.5\textwidth]{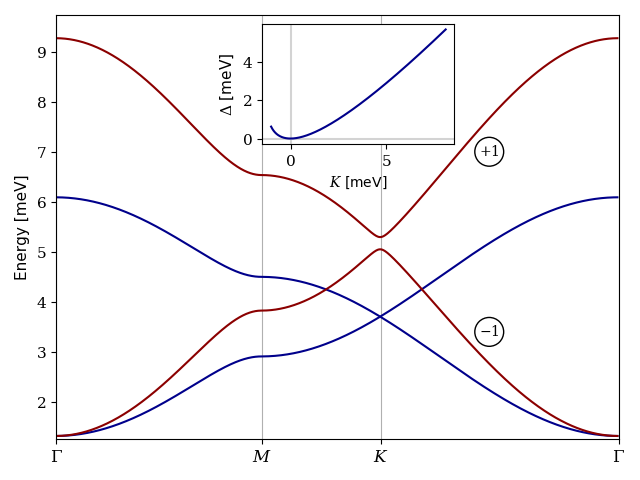}
        \caption{Energy spectrum of magnons within the first Brillouin zone. The blue line corresponds to the case $J=0.53$, $A=0.44$ and $K=0$. It can be see that there is no gap at the $\pmb{K}$-point. There is a gap at the $\Gamma$ point that arises from the anisotropy contribution\cite{Lado_2017}. On the other hand, the inclusion of the Kitaev interaction, $K=2J$ displayed in the red line, displays a gap opening at the $\pmb{K}$-point revealing a non-trivial topology. The circles next to each bands correspond to the associated Chern numbers. These are calculated according to \cite{Murakami}. The $\boldsymbol{K}$-point gap is calculated as a function of the Kitaev interaction strength in the inset.}
         \label{fig: energy spectrum}
\end{figure}

We plot the energy spectrum, across the first Brillouin zone in Fig. \ref{fig: energy spectrum}. The vanishing energy Goldstone mode at the $\mathbf{\Gamma}$ point is lifted to $\Delta_{\mathbf{\Gamma}}$ by the inclusion of the anisotropy term, $A$.
We can see clearly that in the case of absence of the Kitaev contribution the spectrum is degenerate at the $\mathbf{K}$ and $\mathbf{K}^\prime$ points. This degeneracy is lifted by the inclusion of the Kitaev term, leading to the opening of a gap of magnitude $\Delta_\mathbf{K}$.

In the vicinity of the $\mathbf{\Gamma}$ point, the low energy band behaves as:

\begin{equation*}
      \epsilon_{-}(\mathbf{k}) =  2AS + \epsilon_0 a_0^2\left(\frac{2 + 2 \mathcal{A} - \mathcal{K}^2}{24(1+\mathcal{A})}\right) |\mathbf{k}|^2
\end{equation*}
we identify the usual structure
$
\epsilon=\Delta_\mathbf{\Gamma}+\rho_\mathbf{\Gamma} \mathbf{k}^2,$ where $\Delta_\mathbf{\Gamma}=2AS$ and $\rho_\mathbf{\Gamma}=\epsilon_0 a_0^2\left(\frac{2 + 2 \mathcal{A} - \mathcal{K}^2}{24(1+\mathcal{A})}\right)$.
 $\Delta_\mathbf{\Gamma}$ correspond to the minimal energy necessary to create a magnon. It turns out to be a fundamental quantity and can be accessed experimentally. It lies between 1 and 9 $meV$\cite{Jin} while ab-initio calculations locate it in the range of 1 $meV$\cite{Joaquin}.
 $\rho_\mathbf{\Gamma}$ is the effective low energy spin stiffness. It is an estimate of how hard it is to introduce a smooth texture in the magnetization field.

 The behavior of the top band at $\boldsymbol{\Gamma}$ point  is:
 \begin{equation*}
 \epsilon_{+}(\mathbf{k})=2\epsilon_0 + \Delta_{\boldsymbol{\Gamma}} - \rho_{\boldsymbol{\Gamma}}' |\mathbf{k}|^2
 \end{equation*}

 which shows that the bandwidth, defined as the energy difference $\epsilon_{+}-\epsilon_{-}$ at $\boldsymbol{\Gamma}$ point, is given by $2\epsilon_0=2S(3J+K)$.

The effective Hamiltonian in the vicinity of $\mathbf{K}$ and $\mathbf{K}^\prime$ is given by:
\begin{equation}
     \mathbf{H}_{\mathbf{K}^{(\prime)}}(\mathbf{q}) = \left( \begin{aligned}
    \mathbf{T}_{\mathbf{q}} && \mathbf{U}^{(\prime)}\\
    \mathbf{U^{(\prime)\dagger}} && \mathbf{T}_{\mathbf{q}}
    \end{aligned}\right)
    \label{HK}
\end{equation}
where the matrices $\mathbf{T_q}$, $\mathbf{U}$ and $\mathbf{U^{\prime}}$ are defined as:
\begin{equation*}
    \mathbf{T}_{\mathbf{q}} = \epsilon_0 \left( \begin{aligned}
    1+\mathcal{A}&& i\kappa\\
    -i\kappa^{*} && 1+\mathcal{A}
    \end{aligned}\right)
\end{equation*}
and $\mathbf{U}$ is written with the aid of the spin-$\frac{1}{2}$ ladder operator, $\mathbf{U}=K\sigma^-$, The matrix at the other valley is $ \mathbf{U}^{\prime} = \mathbf{U}^{\dagger}$. Here we have defined $\mathbf{q}= \mathbf{k} - \mathbf{K^{(\prime)}}$ and $\kappa= a_0(q_{x} + iq_{y})/(2\sqrt{3})$, with  $a_0$ being the lattice constant. In the definition of $\mathbf{U}$ we have dropped linear terms in $\mathbf{q}$ under the assumtion of small Kitaev parameter. Energies around $\mathbf{K}$ and $\mathbf{K'}$ take the form:
$$
\epsilon^\pm(\mathbf{q})=E_\mathbf{K}\pm\frac{\Delta_\mathbf{K}}{2} \pm \rho^{\pm}_\mathbf{K} |\mathbf{q}|^2
$$

with $E_{\mathbf{K}}=\frac{\epsilon_0}{2}(1+\mathcal{A}+\sqrt{(1+\mathcal{A})^2-\mathcal{K}^2})$ and $\Delta_{\mathbf{K}}= \epsilon_0 (1+\mathcal{A} - \sqrt{(1+\mathcal{A})^2-\mathcal{K}^2})$.

All those features are in agreement with \cite{Joaquin} which can be used to adjust our parameters. We find: $J\sim 0.53$meV, $K\sim 4.07$meV and $A\sim 0.44$meV, in same range as \cite{Lee}.

The band structure found by these method reveals a non-trivial topological structure, which is present both in the full model \eqref{Kitaev_Hk} and even in the minimal model of equation \eqref{HK} . This is in agreement with the results of \cite{Joshi} for a similar geometrical construction.
The Chern numbers of the bands, calculated according to \cite{Murakami}, are displayed in next to each band. The Chern number of the j-th energy band is given by:
$
    C_j = i \frac{\epsilon_{\mu\nu}}{2\pi} \int_{BZ} \text{d}^2k ~Tr\left(\left(1-P_j\right)\left(\partial_{k_\mu}P_j\right)\left(\partial_{k_\nu}P_j\right)\right).
$
The integrand of the Chern number is called the Berry curvature, $\mathbf{\Omega}^j_\mathbf{k}$, and $P_j$ are the projection operators, which are defined as: $
    P_j = T_{\pmb k} \Gamma_j \sigma_3 T_{\pmb k}^\dagger \sigma_3.
$ Where we have that $T_{\pmb k}$ is the transformation matrix obtained by Bogoliubov's algorithm\cite{Colpa}, $\sigma_3$ is the paraunitary matrix and $\Gamma_j$ is a $(2N,2N)$ matrix where every element is 0 except for the j-th diagonal component, where it has a value of $1$.

It is also important to note that the value of the Chern numbers does not revert its sign when $K$ passes from a positive value to a negative one. Therefore an interface between samples with different signs of K would not host topological states as there is no change in the Chern number between the regions.  This is because a chirality is already fixed  when we chose the $\pmb z$ as the quantization axis.  To change the sign of the Chern number we must change and revert the quantization axis.
From this fact we expect  magnetic domain walls on $\text{CrI}_3$ to act effectively as topologically protected waveguides.

The starting Hamiltonian (\ref{Kitaev_H}) displays complete time reversal symmetry.
It is only after its spontaneous breaking that we can expect a non time reversal symmetry (TRS) Hamiltonian for the spin wave branch of excitations. It can be shown that performing TRS is equivalent to change the quantization axis from $\pmb{z}$ to $-\pmb{z}$.
Performing Holstein-Primakoff's transformation around the reversed axis leads to the complex conjugation of the coefficients of equation \ref{Kitaev_Hk}, followed by a $\mathbf{k}\rightarrow-\mathbf{k}$ transformation. The Hamiltonian \eqref{Kitaev_Hk} would be invariant under TRS if the coefficient $\beta_a$ is real. When Kitaev's parameter $K$ is turned on, we obtain $\mathcal{J}_a^{xy} \neq 0$ .  This makes $\beta_a$ complex, so TRS is broken in our Hamiltonian.

It is important to emphasize that the TRS breaking takes place through an anomalous $A-B$ nearest neighbor coupling in contrast to the normal $A-A$ next-nearest-neighbors proposed by \cite{Owerre_2016,Chen, Kim}.

\begin{figure}
        \centering
        \includegraphics[width=0.45\textwidth]{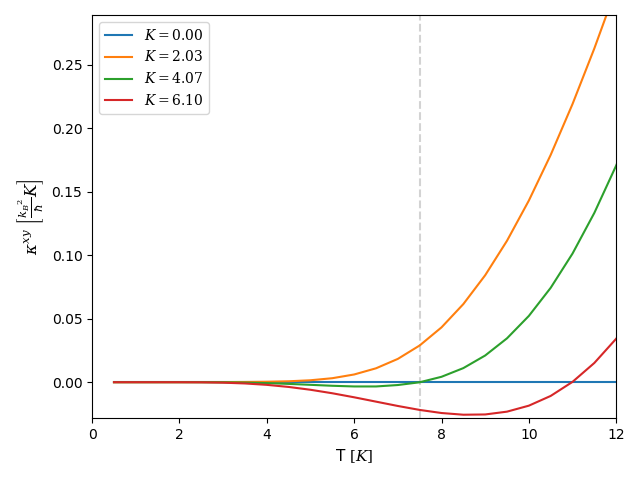}
        \caption{Thermal Hall conductivity $\kappa^{xy}$ vs temperature $T$.  It can be seen that the conductivity changes of sign at different temperatures for different values of $K$.  We highlight that for $K=4.07$meV, the thermal conductivity changes of sign for $T=7.50$K, which is highlited by the grey dotted line.}
        \label{fig: Kappa_xy}
    \end{figure}

{\em Magnon Hall effect.-} The magnon Hall effect corresponds to a transverse magnon-based heat current in response to a thermal gradient. First discovered in the ferromagnetic insulator $Lu_2V_2O_7$ \cite{Onose}, its explanation is understood in terms of magnon Berry's phases~\cite{Katsura,Matsumoto2,Matsumoto,Mook2,Ideue}. The intrinsic contribution associated with the transverse thermal conductivity is written in terms of the Berry curvature as follows~\cite{Matsumoto}:
\begin{equation}
    \kappa^{xy}=-\frac{k^2_B T}{(2\pi)^2\hbar}\sum_n\int_{BZ}\text{d}^2k\, c_2(\rho_n)\mathbf{\Omega}^n_\mathbf{k}
\end{equation}
where $\rho_n=n_B(\epsilon_n(\mathbf{k}))$, $n_B$ being the Bose distribution function, and $c_2(\rho)=\int^\rho_0 (\log(1+t^{-1}))^2{\rm d}t$. In Fig.\ref{fig: Kappa_xy} we show the result for different values of $K$. We note that, for $K=0$, there is no intrinsic contribution to magnon Hall effect. On the other hand, $K\neq0$ leads to a non-vanishing contribution. We highlight the change in sign that the thermal Hall conductivity undergoes, that has already been reported to occur in other materials \cite{Grissonnanche, kasahara2018}.  Due to the monotonically decreasing behavior of $c_2$, the lower band will dominate the sign of the conductivity.  At low temperatures, the function $c_2$ is relevant only in the vicinity of the $\mathbf{\Gamma}$-point, while at higher temperatures it acquires a contribution from the $\mathbf{K}$'s points, where the Berry curvature has the opposite sign, thus explaining the change of sign in the conductivity.

{\em Nanoribbons.-} We now proceed to study the edge states in nanoribbons. This issue has been addressed extensively for magnon topological insulators based, for example, upon the Dzyalonshinskii-Moriya interaction in skyrmion crystals \cite{Rold_n_Molina_2016} and kagome lattices\cite{Mook}. Recently it has been proposed to use edge states to implement topological magnon amplification\cite{Malz}.

\begin{figure}
        \centering
        \includegraphics[width=0.48\textwidth]{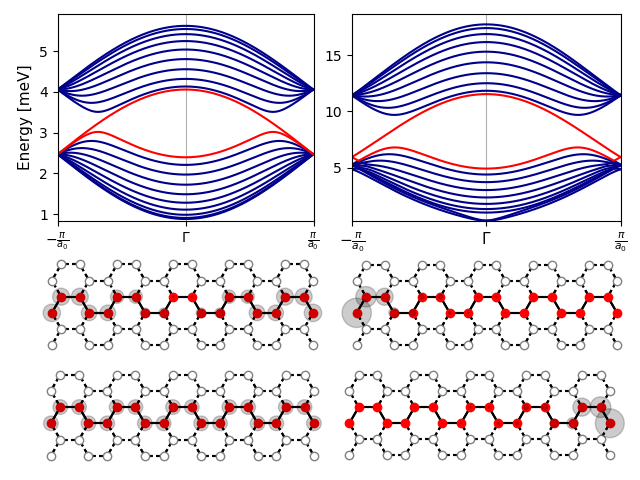}
        \caption{Magnonic energy bands with $N=10$ unit cells zig-zag nanoribbon. \textbf{Left panel}: Situation with $K=0$. We can see a clear resemblance with those of graphene nanoribbons\cite{foa}. The otherwise flat bands in the mid-gap area, rendered in red, are distorted due to the on-site energy discrepancy at the edge. The plots below represent the magnitude of the probabilities (grey circles with increasing radius for larger probabilities) for the mid-gap states at $k^*=0.99\pi/a_0$, represented by the radii of grey circles for the upper central band (top) and the lower central bands  (bottom). The magnon transport is not chiral.
        \textbf{Right panel}: Situation with $K=4.07$meV. The bulk bands preserve their basic shape but the band-width is amplified. The degeneracy of the mid-gap bands, highlighted with red, toward the edge of the Brillouin zone gets lifted. The lower panels describe the probability density of the states at $k^*$ for the upper mid-gap band (top) and the lower mid-gap bands  (bottom). We can se clear evidence of the localization of these states at the geometrical edges of the ribbons.}
        \label{fig:nanoribbon}
    \end{figure}

Despite the starking similarities between the magnon Hamiltonian in absence of the Kitaev term and the usual tight-binding model for graphene~\cite{foa}, there is a subtle difference that becomes relevant in the case of edges, vacancies and similar defects. The diagonal contributions arise from exchange and are, therefore, dependent of the number of neighbors of each site. In this way, the local energy of the sites at the edge is different from the ones in the bulk. As we will see this changes the edge states that become different from their graphene counterparts~\cite{foa}. This can be appreciated in the left panel of Fig.~\ref{fig:nanoribbon}, where the ungapped dispersion in absence of the Kitaev term is displayed. As a result of the shift of the onsite energies at the edge, the flat band typically expected for zig-zag ribbons is now bent toward the lower energy bands.  Interestingly, when the Kitaev term is turned on, as in the right panel of Fig.~\ref{fig:nanoribbon}, the bulk system acquired a gap which is bridged by the states marked in red. As shown in the lower panel, these states are localized around the sample's edges.

\textit{Conclusions.-} In this paper, we have investigated the magnon spectrum of Heisenberg-Kitaev magnets, such as $\text{CrI}_3$.
We have performed a spin wave analysis, based upon a Holstein-Primakoff representation around the out-of-plane preferred direction.
We have found that the Kitaev term propagates the time-reversal symmetry breaking into the magnon sector. This is done through an anomalous nearest-neighbor contribution. That, in turn, leads to topological effects such as a gap in the Dirac point and edge states moving freely along domain walls and edges of the system. We expect that these discoveries will provide a handy tool for magnon based technologies. For example, the topologically protected states propagating at its edges or along domain walls can be used as an efficient method of magnon communication. Additionally the topological states can display thermal Hall effect as shown in \cite{Katsura,Matsumoto,Mook2,Ideue}.
By comparing our results with the all-electron calculations of \cite{Joaquin} we were able to provide early estimates of the magnetic constants of $\text{CrI}_3$.

\textit{Acknowledgments} - A.S.N. thanks Joaquin Fern\'andez-Rossier for helpful comments. The authors thanks Fondecyt Regular 1190324,  N. V.-S. thanks Fondecyt Postdoctorado Nº 3190264. A.S.N., R.J.-U. and E.A. thank to  Financiamiento Basal  para  Centros  Cient\'ificos  y  Tecnol\'ogicos  de  Excelencia FB 0807.

\nocite{*}

\bibliographystyle{apsrev4-1_title}
\bibliography{apssamp}% Produces the bibliography via BibTeX.

\end{document}